\documentclass[useAMS,usenatbib]{mn2e}
\def\aap{AA}

\def\apjl{ApJL}

\def\mnras{MNRAS}
\def\apj{ApJ}
\def\apjs{ApJS}

\def\aj{AJ}

\def\procspie{Proc.~SPIE}

\def\ase{{\prime\prime}}
\def\Jxy{{WGD2038-4008}~}
\def\Jxytwo{{WGD2021-4115}~}

\usepackage{graphicx}
\usepackage{float}
\usepackage{amssymb}
\usepackage{amsfonts}
\usepackage{amsmath} 
\usepackage{color}
\usepackage[switch,pagewise]{lineno}

\def\aaemail{\tt aagnello@eso.org}
\def\eso{European Southern Observatory, Karl-Schwarzschild-Strasse 2, 85748 Garching bei M{\"u}nchen, DE 0000-0001-9775-0331}
\def\valpo{Instituto de Física y Astronomía, Universidad de Valparaíso, Avda. Gran Breta{\~n}a 1111, Playa Ancha, Valparaíso 2360102, Chile }
\def\abello{Departamento de Ciencias Fisicas, Universidad Andres Bello Fernandez Concha 700, Las Condes, Santiago, Chile}
\def\swin{Centre for Astrophysics and Supercomputing, Swinburne University of Technology, PO Box 218, Hawthorn, VIC 3122, Australia}
\def\ucla{Department of Physics and Astronomy, PAB, 430 Portola Plaza, Box 951547, Los Angeles, CA 90095-1547, USA}
\def\kipac{Kavli Institute for Particle Astrophysics and Cosmology, Stanford University, 452 Lomita Mall, Stanford, CA 94305, USA}

\def\mpa{Max-Planck-Institut für Astrophysik, Karl-Schwarzschild-Str. 1, D-85741 Garching, Germany}
\def\ucd{Department of Physics, University of California Davis, 1 Shields Avenue, Davis, CA 95616, USA}

\def\ioa{Institute of Astronomy, Madingley Road, Cambridge CB3 0HA, UK}
\def\kavli{Kavli Institute for Cosmology, University of Cambridge, Madingley Road, Cambridge CB3 0HA, UK}

\def\braza{Laborat\'orio Interinstitucional de e-Astronomia - LIneA, Rua Gal. Jos\'e Cristino 77, Rio de Janeiro, RJ - 20921-400, Brazil}
\def\brazb{Observat\'orio Nacional, Rua Gal. Jos\'e Cristino 77, Rio de Janeiro, RJ - 20921-400, Brazil}
\def\brazc{Departamento de F\'{\i}sica Matem\'atica,  Instituto de F\'{\i}sica, Universidade de S\~ao Paulo,  CP 66318, CEP 05314-970, S\~ao Paulo, SP,  Brazil}
\def\brazd{Instituto de F\'\i sica, UFRGS, Caixa Postal 15051, Porto Alegre, RS - 91501-970, Brazil}
\def\ipmu{Kavli IPMU (WPI), UTIAS, The University of Tokyo, Kashiwa, Chiba 277-8583, Japan}
\def\mit{MIT Kavli Institute for Astrophysics and Space Research, 37-664G, 77 Massachusetts Avenue, Cambridge, MA 02139}
\def\fnal{Fermi National Accelerator Laboratory, Batavia, IL 60510}
\def\epfl{Laboratoire d'Astrophysique, Ecole Polytechnique F\'ed\'erale de Lausanne (EPFL), Observatoire de Sauverny, CH-1290 Versoix, Switzerland}
\def\hyder{Department of Physics, IIT Hyderabad, Kandi, Telangana 502285, India}
\def\slac{Kavli Institute for Particle Astrophysics and Cosmology, Stanford University, 452 Lomita Mall, Stanford, CA 94035, USA}
\def\illa{Department of Astronomy, University of Illinois, 1002 W. Green Street, Urbana, IL 61801, USA}
\def\illb{National Center for Supercomputing Applications, 1205 West Clark St., Urbana, IL 61801, USA}
\def\espa{Institut de Ci\`encies de l'Espai, IEEC-CSIC, Campus UAB, Carrer de Can Magrans, s/n,  08193 Bellaterra, Barcelona, Spain}

\def\cnrs{CNRS, UMR 7095, Institut d'Astrophysique de Paris, F-75014, Paris, France}

\def\espc{Centro de Investigaciones Energ\'eticas, Medioambientales y Tecnol\'ogicas (CIEMAT), Madrid, Spain}
\def\espd{Instituto de Fisica Teorica UAM/CSIC, Universidad Autonoma de Madrid, 28049 Madrid, Spain}
\def\eth{Institute for Astronomy, Department of Physics, ETH Zurich, Wolfgang-Pauli-Strasse 27, 8093, Zurich, Switzerland}

\def\penns{Department of Physics and Astronomy, University of Pennsylvania, Philadelphia, PA 19104, USA}
\def\jpl{Jet Propulsion Laboratory, California Institute of Technology, 4800 Oak Grove Dr., Pasadena, CA 91109, USA}
\def\calt{Department of Physics, California Institute of Technology, Pasadena, CA 91125, USA}
\def\carn{Observatories of the Carnegie Institution of Washington, 813 Santa Barbara St., Pasadena, CA 91101, USA}
\def\mich{Department of Physics, University of Michigan, Ann Arbor, MI 48109, USA}
\def\micha{Department of Astronomy, University of Michigan, Ann Arbor, MI 48109, USA}

\def\ohioa{Center for Cosmology and Astro-Particle Physics, The Ohio State University, Columbus, OH 43210, USA}
\def\ohiob{Department of Physics, The Ohio State University, Columbus, OH 43210, USA}
\def\aao{Australian Astronomical Observatory, North Ryde, NSW 2113, Australia}

\def\slac{SLAC National Accelerator Laboratory, Menlo Park, CA 94025, USA}
\def\sorb{Sorbonne Universit\'es, UPMC Univ Paris 06, UMR 7095, Institut d'Astrophysique de Paris, F-75014, Paris, France}

\def\stanf{Department of Physics, Stanford University, 382 Via Pueblo Mall, Stanford, CA 94305, USA}
\def\ctio{Cerro Tololo Inter-American Observatory, National Optical Astronomy Observatory, Casilla 603, La Serena, Chile}
\def\ucl{Department of Physics \& Astronomy, University College London, Gower Street, London, WC1E 6BT, UK}
\def\cnrs{CNRS, UMR 7095, Institut d'Astrophysique de Paris, F-75014, Paris, France}
\def\sorb{Sorbonne Universit\'es, UPMC Univ Paris 06, UMR 7095, Institut d'Astrophysique de Paris, F-75014, Paris, France}

\def\nick{Staples High School, Westport CT}
\def\ctio{Cerro Tololo Inter-American Observatory, National Optical Astronomy Observatory, Casilla 603, La Serena, Chile}
\def\ucl{Department of Physics \& Astronomy, University College London, Gower Street, London, WC1E 6BT, UK}
\def\rhodes{Department of Physics and Electronics, Rhodes University, PO Box 94, Grahamstown, 6140, South Africa}

\def\wash{Astronomy Department, University of Washington, Box 351580, Seattle, WA 98195, USA}

\def\ifae{Institut de F\'{\i}sica d'Altes Energies (IFAE), The Barcelona Institute of Science and Technology, Campus UAB, 08193 Bellaterra (Barcelona) Spain}
\def\south{School of Physics and Astronomy, University of Southampton,  Southampton, SO17 1BJ, UK}

\def\oak{"Computer Science and Mathematics Division, Oak Ridge National Laboratory, Oak Ridge, TN 37831}
\def\hyder{Department of Physics, IIT Hyderabad, Kandi, Telangana 502285, India}
\def\camp{Instituto de F\'isica Gleb Wataghin, Universidade Estadual de Campinas, 13083-859, Campinas, SP, Brazil}
\def\aaemail{\tt aagnello@eso.org}
\def\eso{European Southern Observatory, Karl-Schwarzschild-Strasse 2, 85748 Garching bei M{\"u}nchen, DE}

\title[WISE-Gaia-DES quasar lenses]{DES meets Gaia: discovery of strongly lensed quasars from a multiplet search} 
\author[A. Agnello et al.]{
  A.~Agnello$^{1}$\thanks{\aaemail, ORCID 0000-0001-9775-0331}, H.~Lin$^{2},$ N.~Kuropatkin$^{2},$ E.~Buckley-Geer$^{2},$ T.~Anguita$^{3},$ P.L.~Schechter$^{4},$\and T.~Morishita$^{5},$ V.~Motta$^{6},$ K. Rojas$^{6},$ T.~Treu$^{5,\dag},$ A.~Amara$^{7},$ M.W.~Auger$^{8},$ F.~Courbin$^{9},$\and  C.D.~Fassnacht$^{10},$  J.~Frieman$^{2},$ A.~More$^{11},$ P.J.~Marshall$^{12},$ R.G.~McMahon$^{8,13},$ G.~Meylan$^{9},$\and S.H.~Suyu$^{14},$ K.~Glazebrook$^{15},$  N.~Morgan$^{16},$  B.~Nord$^{2},$ T. M. C.~Abbott$^{17},$ F.~B.~Abdalla$^{18,19}$\and J.~Annis$^{2},$ K.~Bechtol$^{2},$ A.~Benoit-L{\'e}vy$^{18,20,21},$
 E.~Bertin$^{20,21},$ R.~A.~Bernstein$^{22},$\and D.~Brooks$^{18},$ D.~L.~Burke$^{23,12},$ A.~Carnero~Rosell$^{24,25},$ J.~Carretero$^{26},$ C.~E.~Cunha$^{12},$\and C.~B.~D'Andrea$^{27},$ L.~N.~da Costa$^{24,25},$ S.~Desai$^{28},$ A.~Drlica-Wagner$^{2},$ T.~F.~Eifler$^{29,30},$\and B.~Flaugher$^{2},$ J.~Garc{\'i}a-Bellido$^{31},$ E.~Gaztanaga$^{32}$ D.~W.~Gerdes$^{33,34},$ D.~Gruen$^{12,35},$\and R.~A.~Gruendl$^{36,37},$ J.~Gschwend$^{24,25}$ G.~Gutierrez$^{2}$ K.~Honscheid$^{38,39},$ D.~J.~James$^{40},$\and K.~Kuehn$^{41},$ O.~Lahav$^{18},$ M.~Lima$^{24,25,42},$ M.~A.~G.~Maia$^{24,25},$ M.~March$^{27},$ F.~Menanteau$^{36,37},$\and R.~Miquel$^{26,43},$ R.~L.~C.~Ogando$^{24,25},$
  A.~A.~Plazas$^{29},$ E.~Sanchez$^{44},$ V.~Scarpine$^{2},$\and R.~Schindler$^{35},$ M.~Schubnell$^{33},$ I.~Sevilla-Noarbe$^{44},$ M.~Smith$^{45},$ M.~Soares-Santos$^{2},$\and F.~Sobreira$^{24,46},$ E.~Suchyta$^{47},$ M.~E.~C.~Swanson$^{37},$ G.~Tarle$^{33},$  D.~Tucker$^{2},$ R.~Wechsler$^{23}$
  \\$^1$\eso\\
  $^2$\fnal\\
  $^3$\abello\\
  $^4$\mit\\
  $^5$\ucla\\
  $^6$\valpo\\
$^{7}-^{51}$ Full list of affiliations at the end of the paper.\\
  $\dag$ Packard Fellow.  \\
  \textit{This paper includes data gathered with the 6.5 meter Magellan Telescopes located at Las Campanas Observatory, Chile.}\\
}

\begin{document}

\voffset-.6in

\date{Accepted . Received }

\pagerange{\pageref{firstpage}--\pageref{lastpage}} 

\maketitle

\label{firstpage}

\begin{abstract}
We report the discovery, spectroscopic confirmation and first lens models of the first two, strongly lensed quasars from a combined search in WISE and Gaia over the DES footprint.
 The four-image lens \Jxy (r.a.=20:38:02.65, dec.=-40:08:14.64) has source- and lens-redshifts $z_{s}=0.777\pm0.001$ and $z_{l}=0.230\pm0.002$ respectively. Its deflector has effective radius $R_{\rm eff}\approx3.4^\ase$, stellar mass $\log(M_{\star}/M_{\odot})=11.64^{+0.20}_{-0.43}$, and shows extended isophotal shape variation. Simple lens models yield Einstein radii $R_{\rm E}=(1.30\pm0.04)^\ase,$ axis ratio $q=0.75\pm0.1$ (compatible with that of the starlight) and considerable shear-ellipticity degeneracies.
 The two-image lens \Jxytwo (r.a.=20:21:39.45, dec.=--41:15:57.11) has $z_{s}=1.390\pm0.001$ and $z_{l}=0.335\pm0.002$, and Einstein radius $R_{\rm E}=(1.1\pm0.1)^\ase,$ but higher-resolution imaging is needed to accurately separate the deflector and faint quasar image.
We also show high-rank candidate doubles selected this way, some of which have been independently identified with different techniques, and discuss a DES+WISE quasar multiplet selection.

\end{abstract}


\section{Introduction}
Strong gravitational lensing by galaxies relies on the alignment of a (typically massive) galaxy with a more distant source. In the case of lensed quasars, this enables multiple lines of investigation, such as: the deflector mass and density profile, in both luminous and dark matter \citep[e.g.][]{bat11,ogu14,sch14}; a super-resolved study of distant sources, thanks to magnification \citep{pen06,din17} and micro-lensing by individual stars in the deflector \citep{slu12,hut15,mot17}; an unbiased census of dark and baryonic substructure around the deflector \citep{mao98,dal02,gil17,hsu17}; and the measurement of cosmological distances, from delays between different light-curves \citep{ref64,suy17}. All these investigations are currently limited by sample size. Homogeneous samples of $20-40$ quasar lenses, with a combination of ancillary data (e.g. spectroscopy, time-delay monitoring, deep and high resolution imaging), would enable substantial progress on all fronts.

In order to assemble large samples of lensed quasars, wide areas of sky must be surveyed to overcome the intrinsic rarity of lenses. In the optical, this is made possible by ground-based surveys like the Sloan Digital Sky Survey \citep{aba09},the Dark Energy Survey \citep[hereafter DES,][]{san10, des16}, the VST-ATLAS \citep{sha15} and Pan-STARRS \citep{cha16}. Various techniques have been developed to find quasar lenses, using morphological and spectroscopic information \citep[as in the SQLS and BQLS,][]{ogu06,ina12,mor16}, image cutout modeling \citep{mor04,sch17}, data mining on catalog magnitudes \citep{agn15,ost17,wil17,agn17}, variability \citep{ber17}, and visual inspection of `blue-near-red' objects \citep{lin17,die17}. In the radio, all-sky lens searches \citep[such as CLASS and JVLA,][]{mye03,kin99} examined radio-loud sources that could be resolved into multiple components by higher-resolution follow-up.

The search strategies depend on the quality of the data and on the specific kind of systems being sought. In fact, depending on the projected separation between multiple quasar images, and on the survey image quality and depth, quasar lenses may be catalogued as multiple sources with similar colours, or blended in one or more extended objects. The Gaia satellite mission \citep{gai16,lin16,vLe17}, with sharp image quality (FWHM$\approx0.2^{\ase}$) and moderate depth ($G=20.7$), can be combined with a near-infrared colour-based search from the WISE mission \citep{wri10} to select quasar lenses as quasar-like objects that are resolved into multiple images. The method has been detailed elsewhere \citep{agn17}, here we illustrate its application to the DES year-3 footprint \citep{die16}.

In this paper, we report the discovery of two previously unknown lensed quasars, WGD2038-4008 (r.a.=20:38:02.65, dec.=-40:08:14.64) and WGD2021-4115 (r.a.=20:21:39.45, dec.=--41:15:57.11), from the joint use of WISE colour selection, Gaia multiplet detection, and DES image inspection.
The same search, besides recovering known lenses, has yielded a list of potential doubles, as well as possible Galactic streams and substructure among the contaminants. Throughout this paper, we adopt the following nomenclature: \textit{objects} denotes everything pre-selected via colour cuts; \textit{targets} are objects that have been skimmed with additional techniques (a multiplet search, in this case); and \textit{candidates} are targets that survive pixel-by-pixel examination, such as image cutout modeling or visual inspection. When a WISE source corresponds to multiple DES or Gaia detections, the one closest the the WISE position is denoted as \textit{primary} and the remaining ones as \textit{secondary}.

This paper is organized as follows. In Section~\ref{sec:quad} we
describe our WISE-Gaia-DES search, the discovery of \Jxy~ and \Jxytwo and other candidates,  and spectroscopic confirmation; multi-component fits to the cutouts and gravitational lens models in Section 3. In Section~\ref{sec:disc} we discuss the results and
future prospects of this method. the paper is briefly summarized in Section~\ref{sec:sum}.
Following previous work, DES $grizY$ magnitudes are given in the AB system and WISE magnitudes are in the Vega system.

\begin{figure*}
 \centering
 \includegraphics[width=0.95\textwidth]{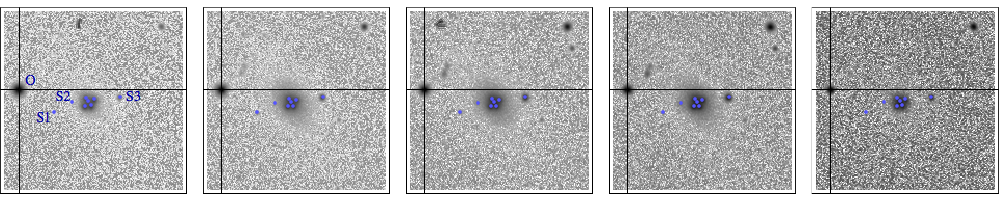}\\
 \includegraphics[width=0.95\textwidth]{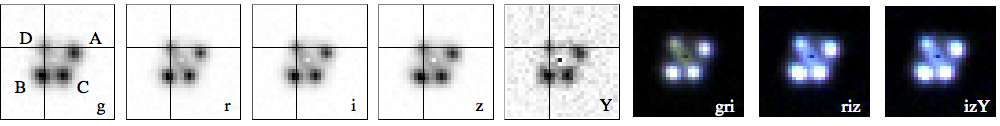}
\caption{{Single-epoch cutouts of WGD2038-4008. \textit{Top:} 45.3$^{\ase}\times$45.3$^{\ase}$ (172px$\times$172px) $grizY$ cutouts, showing the dominant `halo' component of the deflector and a nearby bright star used to fit for the PSF; bullets mark the locations of the objects considered in the cutout modeling, including three sources marked as S1,~S2,~S3. \textit{Bottom:} 7.9$^{\ase}\times$7.9$^{\ase}$ (30px$\times$30px) $grizY$ cutouts and colour-composites, after subtracting the main `halo' profile of the deflector; a flattened bulge, approximately as bright as the faintest quasar image, is visible in the $gri$ cutouts. Images A,~B,~D,~C are marked following the expected ordering in arrival times; the lens galaxy G is in the center.
 }}
\label{fig:WGD2038}
\end{figure*}
\begin{figure*}
 \centering
 \includegraphics[width=0.95\textwidth]{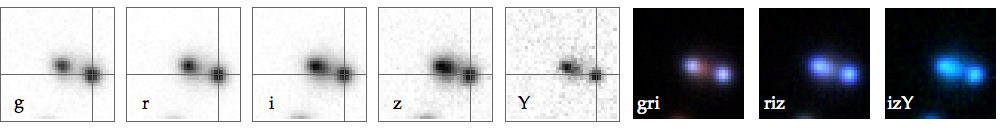}
\caption{{Single-epoch cutouts of WGD2021-4115: 7.9$^{\ase}\times$7.9$^{\ase}$ (30px$\times$30px) $grizY$ cutouts and colour-composites. Axes intersect at image A (the farthest and with shortest arrival time); the counter-image B is blended with the lens G. Due to image quality and B-G proximitiy, uncertainties on astrometry and photometry in WGD2021-4115 are systematics-dominated.
 }}
\label{fig:WGD2021}
\end{figure*}
\section{The Wise-Gaia-DES search, and discovery of lensed quasars}
\label{sec:quad}

Multiple data-mining techniques have been applied to the search of lensed quasars in the DES, over the last two years. Some objects were flagged independently within different searches (see tab.~\ref{tab:candtab}). WGD2038-4008 and WGD2021-4115 (shown in figures~\ref{fig:WGD2038}, \ref{fig:WGD2021}) were discovered using a novel search combining WISE, Gaia and DES.
General properties of this search strategy and numbers for other survey footprints are given by \citet{agn17}, here we summarize the main steps: first, objects are first selected based on their WISE colours; then, those corresponding to multiple-source detections (i.e. `multiplets') in Gaia are retained as targets; finally, DES image cutouts of targets are visually inspected to identify lens candidates. A variation on this theme will be illustrated in Section~\ref{sect:desmul}.

\subsection{WISE-Gaia-DES search}
\begin{figure*}
 \centering
 \includegraphics[width=0.45\textwidth]{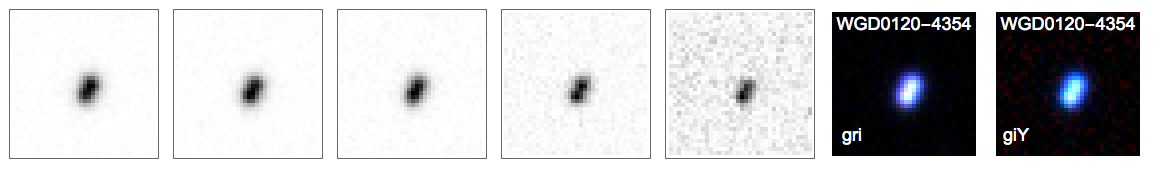}
 \includegraphics[width=0.45\textwidth]{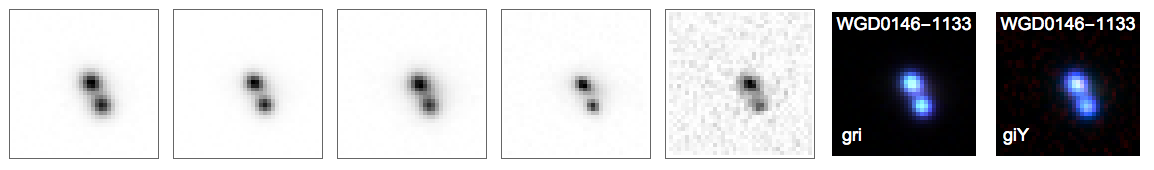}\\
 \includegraphics[width=0.45\textwidth]{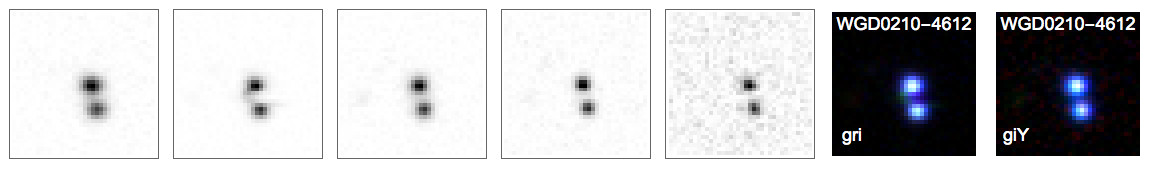}
 \includegraphics[width=0.45\textwidth]{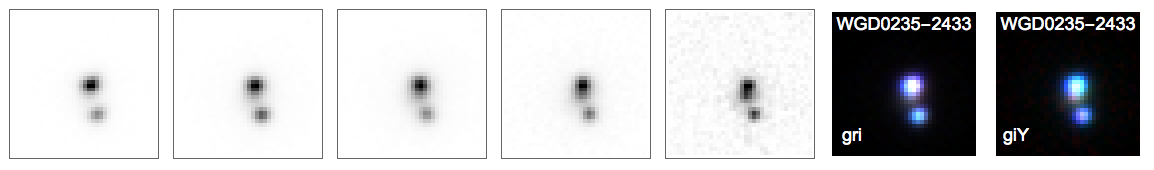}\\
 \includegraphics[width=0.45\textwidth]{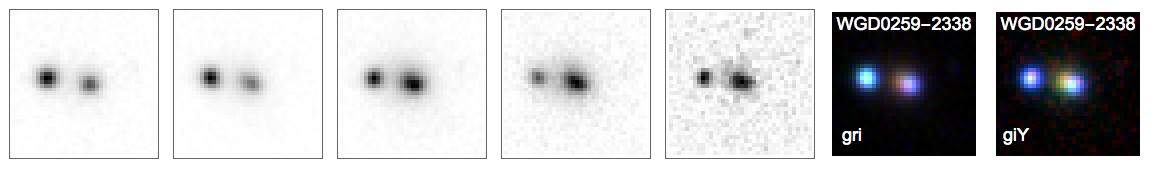}
 \includegraphics[width=0.45\textwidth]{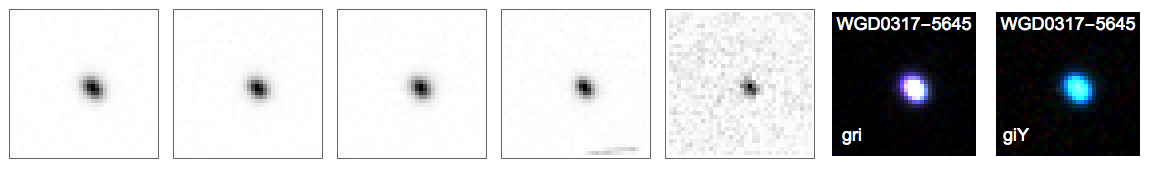}\\
 \includegraphics[width=0.45\textwidth]{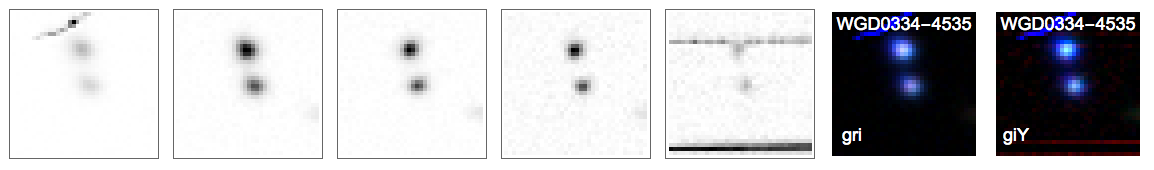}
 \includegraphics[width=0.45\textwidth]{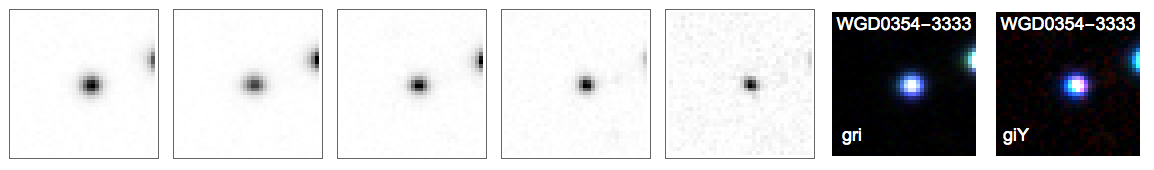}\\
 \includegraphics[width=0.45\textwidth]{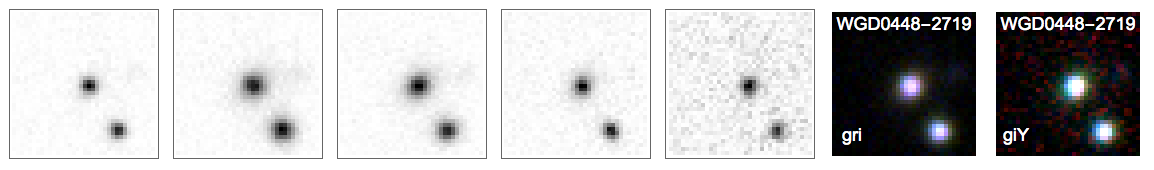}
 \includegraphics[width=0.45\textwidth]{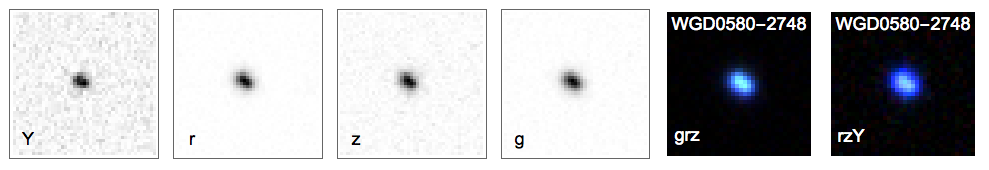}\\
 \includegraphics[width=0.45\textwidth]{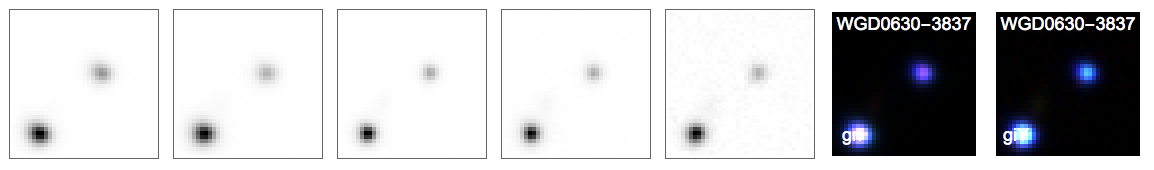}
 \includegraphics[width=0.45\textwidth]{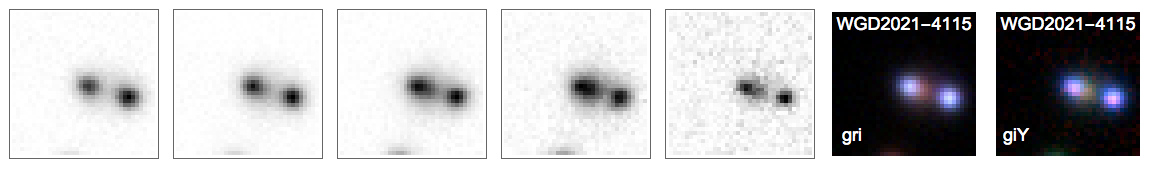}\\
 \includegraphics[width=0.45\textwidth]{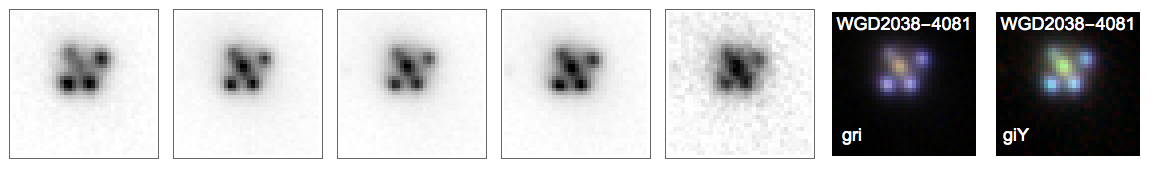}
 \includegraphics[width=0.45\textwidth]{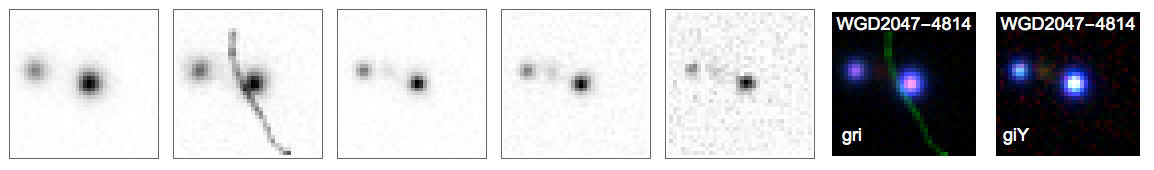}\\
 \includegraphics[width=0.45\textwidth]{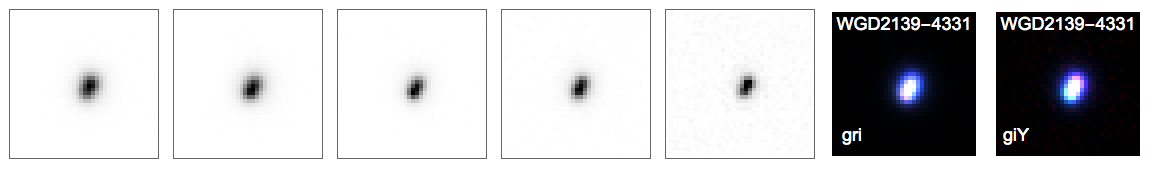}
 \includegraphics[width=0.45\textwidth]{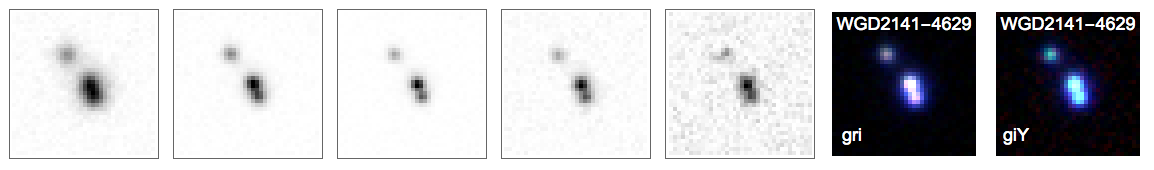}\\
 \includegraphics[width=0.45\textwidth]{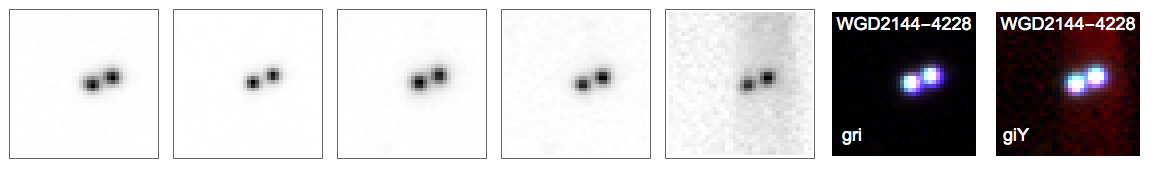}
 \includegraphics[width=0.45\textwidth]{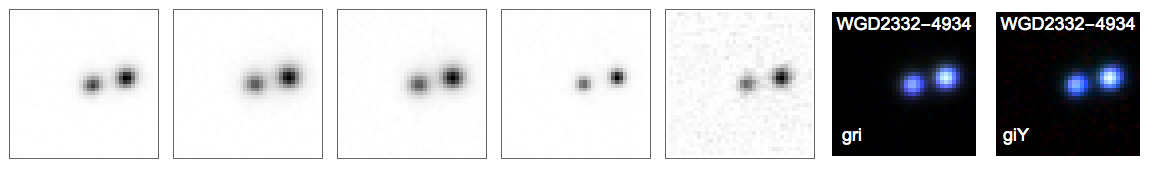}\\
 \includegraphics[width=0.45\textwidth]{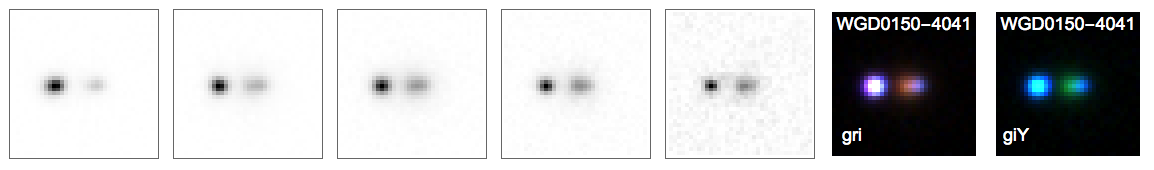}
 \includegraphics[width=0.45\textwidth]{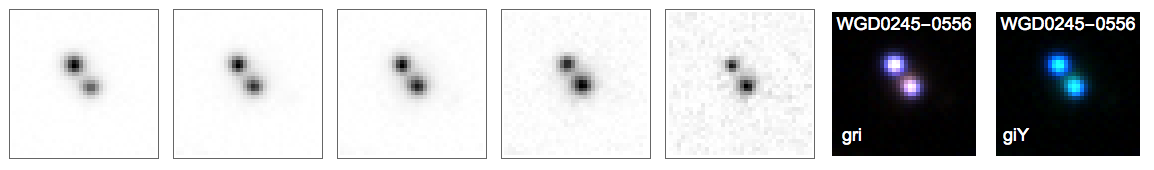}
\caption{{DES single-epoch cutouts of high-grade, multiplet-selected candidates that are not already known lenses. The last two (WGD0150-4041, WGD0245-0556) have been found among DES multiplets, and correspond to singlets in Gaia-DR1. Three multiplets (WGD0259-2338, WGD0146-1133, WGD0235-2433) have also been identified as WISE-Gaia multiplets in the ATLAS-DR3 footprint, where they have coverage in just two bands.
 }}
\label{fig:multcands}
\end{figure*}

The initial sample consisted of 173048 objects, selected in a footprint slightly larger than the DES-Y3 one, satisfying\footnote{Here W`X' and $\delta$W`X' (`X'=1,2,3) refer to the magnitudes \texttt{w`X'mpro} and respective uncertainties \texttt{w`X'sigmpro} from the WISE catalogue.}
\begin{eqnarray}
\nonumber W1-W2>0.55,\  2.2<W2-W3<3.8\\
\nonumber W1<17.0, W2<15.4, W3<11.6, \\
\nonumber \delta W1<0.25,\ \delta W2<0.3,\ \delta W3<0.35,\\
 W2-W3<\mathrm{max}[2.7;\, 3.15+1.5(W1-W2-1.075)]\ .
\label{eq:wisecols}
\end{eqnarray}
A cross-match with Gaia yielded 107076 sources, of which 2656 are unique multiplets - i.e. WISE sources corresponding to more than one non-duplicate Gaia detection within 6$^\ase$ from the corresponding WISE position. A significant subset of these ($\approx 2000$) lie within the DES-Y3 footprint. This target list was narrowed down to a sample of 53 candidates by simple visual inspection of their DES images. Among these are some recently discovered lenses, which had been previously identified with other techniques \citep[Anguita et al. in prep.]{agn15b,lin17}. The WFI2033-4723 quad \citep{mor04} was identified as well, whereas WFI2026-4536 was excluded by the WISE colour pre-selection. In order to prioritize follow-up, candidates can be further graded based on how closely they resemble lenses, based mostly on two criteria: (\textsc{i}) the presence of two or more blue, compact sources with consistent colours across multiple bands; (\textsc{ii}) the presence of a (typically red) galaxy between the putative quasar images. Acceptable `new' candidates identified within this search are shown in Figure~\ref{fig:multcands} and listed in Table~\ref{tab:candtab}.

Some veritable multiplets may have been classified as duplicates in the Gaia-DR1, and so were not flagged by this search. A complementary search, using DES multiplets instead of Gaia-DR1, is outlined in Section~\ref{sect:desmul}. It recognized some WISE-Gaia multiplets, and gave at least two additional candidates (tab.~\ref{tab:candtab}) as WISE-DES multiplets that would correspond to WISE-Gaia singlets.

\begin{table} 
\centering
\begin{tabular}{lc|ccc|}
\hline
 name  &  r.a. & dec.  & $i$ (mag) \\
\hline
WGD0120-4354$^{(d)}$  &  20.026573 & -43.911281  & 19.09  \\
WGD0146-1133$^{(a,d)}$  &  26.636987 & -11.560821  & 17.48  \\
WGD0210-4612  &  32.728207 & -46.214376  & 19.49  \\
WGD0235-2433$^{(a)}$  &  38.864257 & -24.553678  & 17.12  \\
WGD0259-2338$^{(a)}$  &  44.889649 & -23.63383  & 18.41  \\
WGD0317-5645$^{(d)}$  &  49.429068 & -56.081904  & 18.74  \\
WGD0334-4535  &  53.672139 & -45.064588  & 19.90  \\
WGD0354-3333  &  58.738739 & -33.560711  & 19.04  \\
WGD0448-2719  &  72.180042 &-27.325524  & 19.22  \\
WGD0508-2748  &  77.013318 &-27.80528  & 19.23  \\
WGD0630-3837  &  90.783347 & -38.628656  & 18.64  \\
WGD2021-4115  &  305.414396 & -41.265865  & 18.51  \\
WGD2038-4008  &  309.511278 & -40.137107  & 16.50  \\
WGD2047-4814$^{(c,d)}$  &  311.862059 & -48.029964  & 19.11  \\
WGD2139-4331  &  324.988092 & -43.517187  & 17.58  \\
WGD2141-4629$^{(d)}$  &  325.453535 & -46.496119  & 18.95  \\
WGD2144-4228  &  316.18399 & -42.481507  & 17.73  \\
WGD2332-4934  &  353.042968 & -49.568522  & 18.98  \\
WGD0150-4041$^{(b)}$  &  27.736921 & -40.695568  & 18.66  \\
WGD0245-0556$^{(b)}$  &  41.356506 & -5.950145  & 18.73  \\
\hline
\end{tabular}
\caption{Concise list of high-grade, multiplet-selected lens candidates that are not already known lenses. The $i-$band (\texttt{wavg{\_}mag{\_}auto}) magnitude refers to the primary DES match of each WISE-detected object, which is often a blend of multiple components. $^{(a)}$ The same search over the ATLAS-DR3 footprint has yielded candidates that also lie in the DES footprint. $^{(b)}$ The last two candidates have been identified among DES multiplets that correspond to Gaia singlets. $^{(c)}$ Long-slit spectroscopy (SOAR-Goodman, PI V.~Motta) excludes WGD2047-48 as a line-of-sight alignment of a narrow-line-galaxy at $z=0.33$ with a red galaxy and a blue star. $^{(d)}$ Some of the WISE-Gaia candidates had also been identified independently, with other techniques, in the STRIDES-2016 campaign (Anguita et al., Treu et al.; in prep.).}
\label{tab:candtab}
\end{table}
\begin{table*} 
\centering
\begin{tabular}{|cc||c|c||c|c|c|c|c|}
\hline
obj. & comp.  & $\delta x$($^{\ase}$) & $\delta y$($^{\ase}$) & $g$ & $r$ & $i$ & $z$ & $Y$\\
 &  & $= -\delta\mathrm{r.a.}\cos(\mathrm{dec.})$ & $\delta \mathrm{dec.}$ &  &  &  &  & \\
\hline
J2038-4008 & A & 0.00$\pm0.04$ & 0.0$\pm0.04$ & 20.22$\pm$0.03 & 19.74$\pm$0.03 & 19.24$\pm$0.03 & 18.67$\pm$0.03 & 18.96$\pm$0.03 \\ 
 & B & $-2.22\pm0.05$ & $-1.59\pm0.05$ & 20.08$\pm$0.04 & 19.45$\pm$0.04 & 18.98$\pm$0.04 & 18.50$\pm$0.05 & 18.71$\pm$0.03\\ 
 & C & $-0.79\pm0.05$ & $-1.54\pm0.06$ & 20.17$\pm$0.04 & 19.50$\pm$0.04 & 19.08$\pm$0.03 & 18.58$\pm$0.03 & 18.66$\pm$0.03 \\ 
 & D & $-2.01\pm0.05$  & $0.27\pm0.07$ & 20.99$\pm$0.05 & 19.91$\pm$0.06 & 19.42$\pm$0.06 & 19.95$\pm$0.05 & 19.24$\pm$0.10 \\ 
 & G & $-1.38\pm0.04$ & $-0.45\pm0.06$ & 19.18$\pm$0.015 & 18.02$\pm$0.016 & 17.54$\pm$0.016 & 17.19$\pm$0.016 & 16.80$\pm$0.015 \\ 
 & S1 & $-9.96\pm0.06$ & $-3.17\pm0.07$ & 24.10$\pm$0.03 & 22.48$\pm$0.03 & 22.12$\pm$0.02& 21.87$\pm$0.03 & 21.82$\pm$0.07\\ 
 & S2 & $-5.57\pm0.15$ & $-0.71\pm0.09$ & [ $27.5\pm0.6$ ] & [ $25.1\pm0.5$ ] & 22.82$\pm$0.08 & 21.93$\pm$0.08 & 21.33$\pm$0.08\\ 
 & S3 & $6.58\pm0.06$ & $0.68\pm0.04$ & 21.82$\pm$0.01 & 20.57$\pm$0.03 & 19.21$\pm$0.02 & 18.79$\pm$0.03 & 18.49$\pm$0.02 \\ 
\hline
J2021-4115 & A & 0.00$\pm0.06$ & 0.0$\pm0.06$ & 19.81$\pm$0.01 & 18.90$\pm$0.01 & 18.73$\pm$0.01 & 18.53$\pm$0.02 & 18.29$\pm$0.01 \\ 
 & B & -2.50$\pm$0.13  &  0.75$\pm$0.13  &  19.80$\pm$0.03  & 18.80$\pm$0.07  &  18.86$\pm$0.10  &  18.60$\pm$0.10  &  18.53$\pm$0.07 \\ 
 & G &  -2.04$\pm$0.13  &  0.46$\pm$0.13  &  ---  &  ---  &  20.2$\pm$0.4  &  19.9$\pm$0.3  &  19.5$\pm$0.3  \\ 
\hline
\end{tabular}
\caption{Positions (relative to image $A$) and magnitudes of the objects in the two lenses, from a joint model of the DES $grizY$ single-epoch images with best image quality. The quasar images are named according to the expected time-delays.
\textit{Top:} photometry and (relative) astrometry of \Jxy. The PSF is fit directly to the bright star in fig.~\ref{fig:WGD2038}, and three nearby compact sources are included in the fit, in order to limit systematic uncertainties on positions and magnitudes. The magnitudes quoted for G are from the main halo component, whose light dominates over the compact bulge (visible in the residuals), which is roughly as bright as image D. Uncertainties on displacements and magnitudes of image D are systematically dominated, due to contamination by G. Magnitudes in square brackets (for source S2) may be significantly affected by noise and contamination by G.
\textit{Bottom:} photometry and (relative) astrometry of \Jxytwo. The PSF is fit to the farthest image A, together with positions and magnitudes of the three components. The cutout image quality, proximity of B and G, and small extent of G render the fitted positions and magnitude significantly uncertain.
}
\label{tab:SEDs}
\end{table*}

While other lens candidates would require spectroscopic follow-up for confirmation, \Jxy stood out due to its distinctive configuration, with four blue point-source images around a luminous red galaxy. Figure~\ref{fig:WGD2038} shows multi-band wide-field cutouts of the system and a close-up, with our adopted naming convention for different components and nearby sources.

\subsection{Spectroscopic confirmation}
Both \Jxy and \Jxytwo were observed with long-slit spectroscopy, to secure their lens and source resdhifts, as part of a campaign to obtain follow-up spectroscopy of the lens galaxy in various quasar lenses. To this aim, IMACS at the 6.5m Walter Baade Telescope at Magellan (Las Campanas) was used, set up in `long' f/4 camera mode, covering the full $4000${\AA}-$12000${\AA} wavelength range. For WGD~2038-4008, the slit was simply oriented North-South and centered on the lens galaxy; since \Jxytwo required proper spectroscopic confirmation, the slit was oriented along the two blue images.

\begin{figure}
 \centering
 \includegraphics[width=0.45\textwidth]{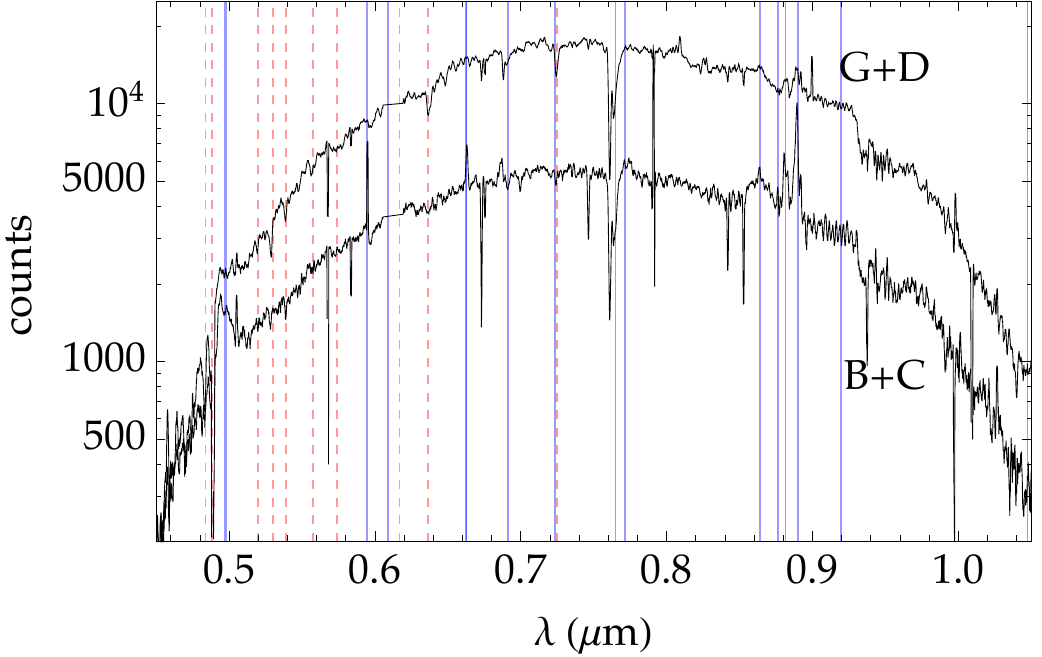}\\
 \includegraphics[width=0.45\textwidth]{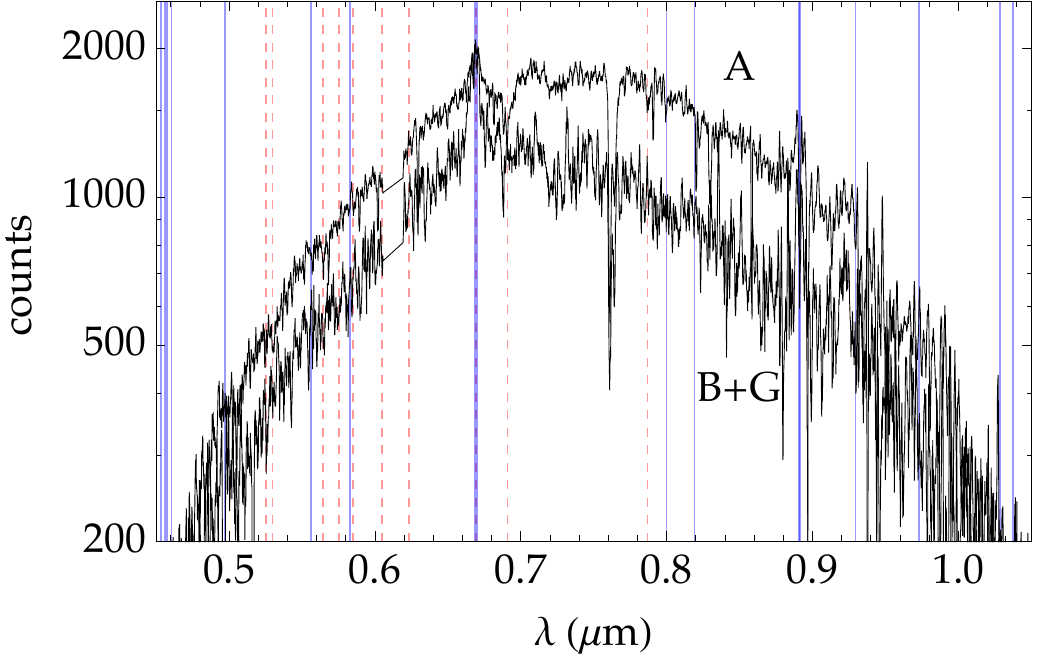}\\
\caption{{Long-slit spectra of \Jxy (\textit{top}) and \Jxytwo (\textit{bottom}) from Magellan-IMACS follow-up.
 }}
\label{fig:specs}
\end{figure}
Figure~\ref{fig:specs} shows the extracted spectra of both systems. From emission lines, the source resdshifts of the quad and double are $z_{s}=0.777\pm0.001$ and $z_{s}=1.390\pm0.001$, respectively. Absorption features, compatible with $z_{l}=0.230\pm0.002$ on the quad and $z_{l}=0.335\pm0.002$ on the double, are also identified.

\section{Cutout models and lens properties}
Simple lens properties can be obtained by modeling the DES image cutouts. In particular, from the $grizY$ survey images we obtained photometry and relative astrometry of multiple components in the two confirmed lenses, and simple lens models based on those.

\subsection{WGD2038-4008: system configuration and gravitational lens models}


WGD2038-4008 has a quite symmetric configuration: two saddle-point images aligned along the major axis of the deflector, and the two minima on a nearly perpendicular line. These components are all blended together by the DES pipeline, but can be disentangled by modeling the DES cutouts as a superposition of four point-sources and a galaxy, for which we choose a single S{\'e}rsic profile \citep{ser68}.

\subsubsection{Photometry and Astrometry of  WGD2038-4008}

We use a nearby, bright point source (indicated as `O' in fig.~1) to directly model the point-spread function (PSF) and the slight offset among different bands. In each band, the PSF is well fit (to within 10\%) by a superposition of three concentric Gaussians with the same axis ratios and orientations, contributing 61\%, 33\% and 6\% of the total flux in the core, wings and outskirts respectively. Residuals from PSF mismatch are due to multipolar features that are not captured by these simple models.

Table~\ref{tab:SEDs} lists the relative positions and $grizY$ magnitudes of the four images (A,B,C,D), deflector (G) and three additional sources (S1, S2, S3). The deflector has a compact bulge and an extended and luminous `halo', which contributes most of the light and is fit by a S{\'e}rsic profile with $R_{\rm eff}=(3.45\pm0.12)^\ase,$ $n_{\rm s}=1.90\pm0.05,$ axis ratio $q=0.77\pm0.01$ and p.a. $\phi_{l}=(41.6\pm1.6)$~deg (E of N). Faint features in blue bands are visible, with isophotes twisting towards 30deg East of North. Contamination by the deflector affects the position and fluxes of image D, whose uncertainties are systematics-dominated.

\begin{table*} 
\centering
\begin{tabular}{lcc|ccccc|cc|}
\hline
obj. & mod.  &  $\theta_{\rm E}$ & $q$  & $\phi_{l}$  & $\phi_{\rm s}$ & $\gamma_{\rm s}$ & $\delta x_{G}$ & $\delta y_{G}$ \\
&   &  ($^\ase$) &    & (deg. E of N)  & (deg. N of W)  &  &  ($^\ase$) &  ($^\ase$)  \\
\hline
J2038-4008  &  mod.(a)  &  1.26$\pm$0.03 & 0.79$\pm$0.05  & 42.0$\pm$7.7  & 25.0$\pm$7.0 & 0.095$\pm$0.025 & -0.04$\pm$0.03 & 0.04$\pm$0.04 \\
 &  mod.(b) &  1.26$\pm$0.05 & $0.64\pm0.22$ & $42.0\pm7.0$  & 18.0$\pm$35.0 & $0.07\pm0.04$ & $-0.06\pm$0.04 & 0.05$\pm$0.04  \\
 &  mod.(c)  &  1.32$\pm0.04$ & 0.44$\pm0.06$  & 32.2$\pm5.05$  & $-75.4\pm50.3$ & $0.02\pm0.02$ & $-0.06\pm0.04$ & $0.05\pm0.04$  \\
 \hline
 J2021-4115  &  mod.(a)  &  1.25$\pm$0.05  & 0.50$\pm$0.08   & 8.0$\pm$4.0   & ---  & [0.0]  & 0.02$\pm$0.06  & 0.00$\pm$0.06  \\
 &  mod.(b) &  1.06$\pm$0.07 & [1.0]  & ---   & -11.0$\pm$5.0  & 0.21$\pm$0.04  & 0.03$\pm$0.06  & 0.02$\pm$0.06   \\
\hline
\end{tabular}
\caption{Lens model parameters: Einstein radius $\theta_{\rm E}=b/\sqrt{q},$ axis ratio $q,$ and p.a. $\phi_{l}$ of the main lens; external shear angle $\phi_{s}$ and amplitude $\gamma_{s};$ and lens center ($\delta x_{G},$ $\delta y_{G}$) with respect to the galaxy center identified in the DES cutouts. Image positions are fit with a SIE for the deflector (G) and external shear to account for possible external contributions. Different lines correspond to different priors adopted on the shape parameters $q,\phi_{l}$ of G (see text).}
\label{tab:lenstab}
\end{table*}

\begin{figure}
 \centering
 \includegraphics[width=0.45\textwidth]{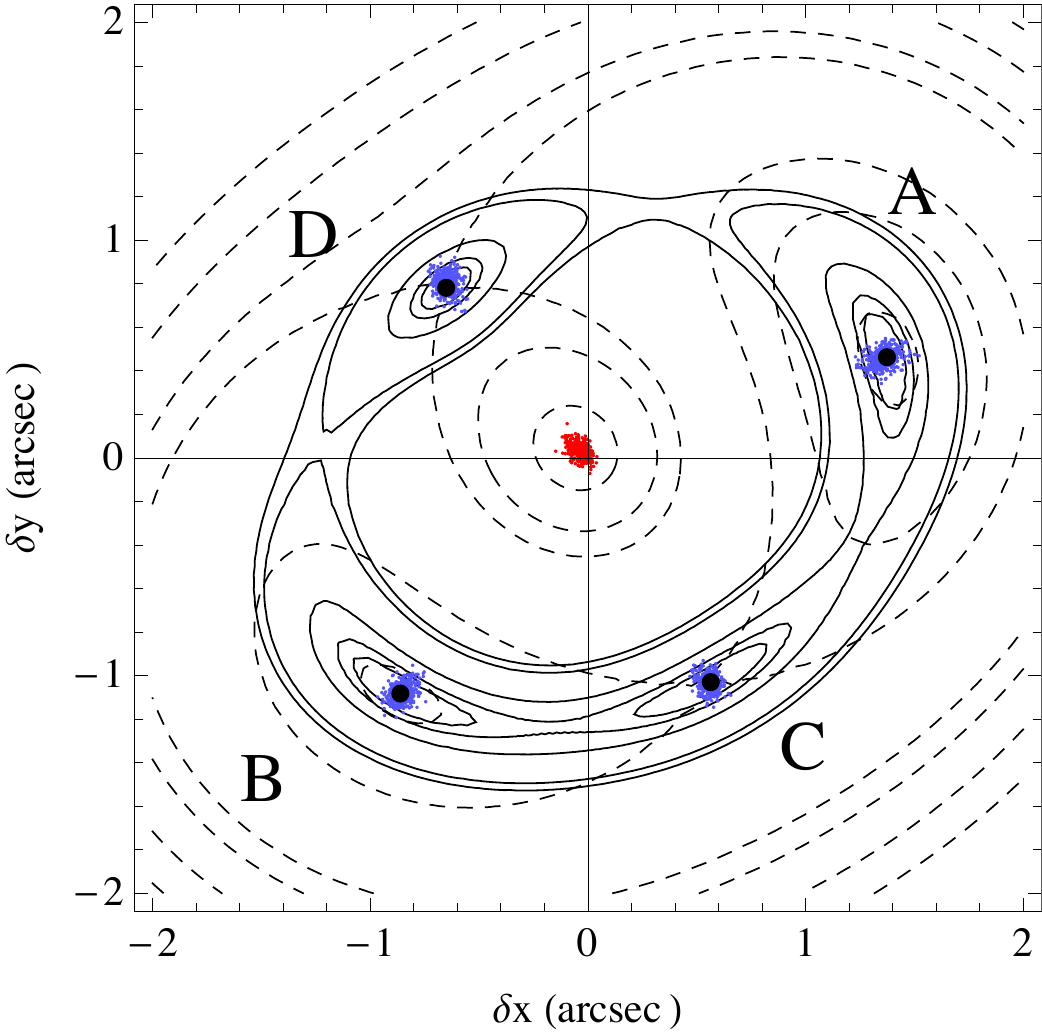}\\
 \includegraphics[width=0.45\textwidth]{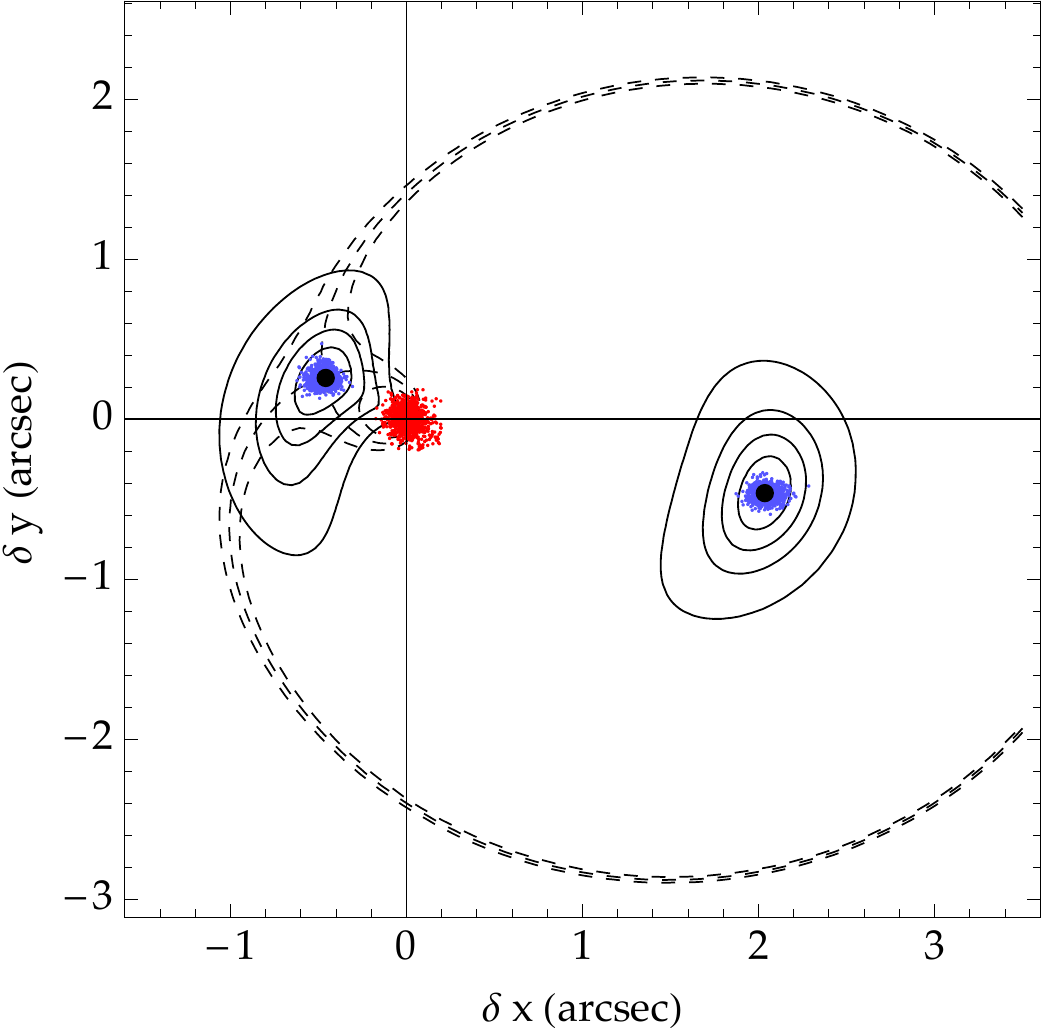}

\caption{{Lens properties from cutout modeling: best-fit SIE models  (see tab.~\ref{tab:lenstab}), with arrival-time contours (dashed) and images of circular source-plane isophotes (full lines), for \Jxy (\textit{top}) and \Jxytwo (\textit{bottom}).
 }}
\label{fig:lensmods}
\end{figure}

\subsubsection{Properties of the deflector galaxy}

From the magnitudes of G as measured from the cutouts, we can infer its stellar mass. To this aim, we use the \textsc{Fast} code \citep{kri09}, with an exponentially-declining star-formation history and a Salpeter stellar IMF. We adopt the nominal magnitude uncertainties ($\delta$m=0.016) from Table~\ref{tab:SEDs}, 
 obtaining $\log_{10}(M_{\star}/M_{\odot})=11.40^{+0.01}_{-0.08}.$
The highly skewed confidence intervals for $M_{*},$ are a consequence of the small (statistical) uncertainties on the measured magnitudes and, consequently, of the steepness of the $4000\AA$ break. With current, broad-band data, different stellar templates are indistinguishable. 

Simple lens models can be fit using the measured image positions as constraints. We do not consider flux-ratio constraints, as they can be easily affected by differential reddening, microlensing and time-delays \citep[e.g.][]{yon99}. We adopt a singular isothermal ellipsoid for the deflector, with deflections
 \begin{eqnarray}
 \alpha_{X}\ =\ -\frac{b}{\sqrt{1-q^{2}}} \arctan\left(\frac{X\sqrt{1-q^{2}}}{\sqrt{q^{2}X^{2}+Y^{2}}}\right)\\
 \alpha_{Y}\ =\ -\frac{b}{\sqrt{1-q^{2}}} {\rm arctanh} \left(\frac{Y\sqrt{1-q^{2}}}{\sqrt{q^{2}X^{2}+Y^{2}}}\right)
 \end{eqnarray}
in a system aligned along the principal axes $(X,Y)$ of G. Here, the deflector is described solely by its lens strength $b$ and axis ratio $q,$ besides its position angle $\phi_{l}.$ We add an external shear component with uniform prior on the amplitude within $0.0<\gamma_{s}<0.14$ and uniform prior of the shear angle $0.0<\phi_{s}<\pi,$ and explore three different choices for the deflector shape parameters: (a) priors on $q$ and $\phi_{l}$ centered on those from cutout modeling, with larger adopted uncertainties of $0.1$ and $10$~deg respectively; (b) $\phi_{l}$ fixed to its value from cutout models, and $q$ free to vary; (c) uniform priors on $q,\phi_{l}.$ In all models, the Einstein radius $\theta_{\rm E}=b/\sqrt{q}$ has a uniform prior between $1.0^\ase$ and $3.0^\ase.$

Table~\ref{tab:lenstab} shows the lens model results.
The Einstein radius is quite robust against the adopted priors, and approximately half the A-B image separation. The lens position angle is always close to those observed from the cutouts, whereas a significant shear-ellipticity degeneracy is present. More accurate lens models should account for the variation in shape and position angle of the observed isophotes (e.g. by using a superposition of ellipsoidal models), and possibly incorporate observations of galaxies in the vicinity of the lens in projection, to independently assess the shear component.

\subsection{WGD2021-4115: system configuration and gravitational lens models}
\Jxytwo consists of two quasar images at either side of a round, faint, red galaxy (fig.~\ref{fig:WGD2021}). We denote the farther image (with shorter arrival time) as A. Its counterimage B lies very close to the deflector galaxy.
The same approach as above has been followed for \Jxytwo, obtaining object positions and $grizY$ magnitudes from the DES image cutouts and performing simple lens models. The astrometry and photometry of the three components is given in Table~\ref{tab:SEDs}. In this case, G is smaller, fainter, and very close to image B, which makes the cutout models highly degenerate. The structural parameters of G cannot be recovered reliably, and its magnitudes have significant uncertainties.

From simple lens models, a high quadrupole component (shear $\gamma_{s}=0.20\pm0.04$ or flattening $q=0.50\pm0.08$) is needed to account for the off-centering of images A,~B with respect to the lens galaxy, still producing only two quasar images due to the highly asymmetric configuration. The Einstein radius ($(1.25\pm0.05)^{\ase}$ from SIE, $(1.06\pm0.07)^{\ase}$ from SIS+XS) is comparable to half the A-B image separation, but due to the significant quadrupole it depends appreciably on the chosen lens model.

\section{Discussion}
\label{sec:disc}

The combination of WISE colour selection and Gaia/DES image quality and depth has resulted in the discovery of the lensed quasar quad \Jxy and a sample of high-grade candidate doubles, among which \Jxytwo was spectrocopically followed up and confirmed. Due to different survey strategy, source-detection pipelines, and depth and image quality, Gaia can detect multiple sources that are unrecognized in DES and \textit{vice versa}. Also, the current performance of the WISE-Gaia multiplet search is limited by the pre-selection cuts in WISE and absence of colour information in Gaia-DR1. These aspects are discussed below.

\begin{figure*}
 \centering
 \includegraphics[width=0.99\textwidth]{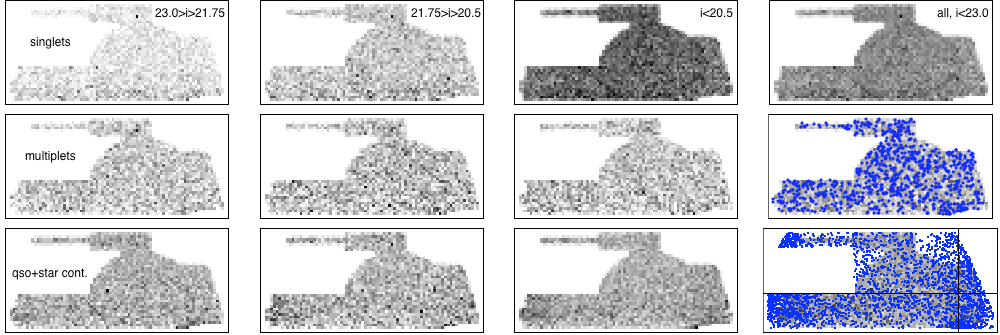}
\caption{{WISE-DES singlets and multiplets. Each row of density maps corresponds to a different macro-class: WISE sources corresponding to DES singlets; WISE sources corresponding to (possibly) extragalactic multiplets; and WISE sources corresponding to multiplets that likely contain a foreground star. In the second line, blue points mark WISE-DES multiplets whose primary has $i<20.5$ (\texttt{wavg{\_}mag{\_}auto}); in the third line, the blue points are the 2656 WISE-Gaia target multiplets. The candidate streams WG1,...,WG4 of \citet{agn17} can be mapped in different magnitude slices; axes intercept at the location of the Globular Cluster NGC 1851.
 }}
\label{fig:tank}
\end{figure*}

\subsection{DES multiplets}
\label{sect:desmul}

The WISE-Gaia search took 173048 input positions, but resulted in fewer (107076) output detections. This is mostly because of the current Gaia-DR1 completeness towards faint magnitudes, and pipeline-specific choices in source detection and deblending. By contrast, a WISE-DES match resulted in 206671 entries, meaning that $\approx50\times10^3$ WISE objects correspond to multiple DES-detected sources within a search radius of $10^\ase.$ This suggests that some interesting objects are missed or mis-classified by Gaia but can be recovered within DES multiplets.

On the other hand, many WISE-DES multiplets can contain very faint objects, mostly galaxies, or line-of-sight alignments of extragalactic objects with foreground stars. Foreground star contamination is already present in the WISE-Gaia selection, as the distribution of multiplets is more abundant towards the Galactic disc, the Large Magellanic Cloud and along four candidate streams \citep[designated as WG1,...,WG4 by][]{agn17}. In a WISE-DES multiplet search, $grizY$ magnitudes can be used to infer which multiplets are most likely contaminants, and which ones are more promising, as one can expect a qso+star alignment to show two well-separated objects, possibly with different colours.

A simple search can then exclude multiplets with separation $\gtrsim3^\ase$ and with at least one point-source, or multiplets whose primary has promising quasar colours ($g-i<0.85$) and a colour difference $\left|\delta(g-i)\right|>0.4$ between primary and secondary detected sources. The separation criterion is chosen because, at large Einstein radii, the lens galaxy should be detectable by the DES pipeline.

If this selection is effective, what remains should be almost uniformly distributed over the DES footprint. For a first assessment of the efficiency of this procedure, and account for footprint coverage, one can simply plot the distribution of different systems.
Figure~\ref{fig:tank} shows the distribution of WISE-DES singlets, WISE-DES multiplets that survive the above cuts in separation and colour difference, and WISE-DES multiplets that should be mostly extragalactic+star alignments. Three distributions are shown, corresponding to singlets/multiplets whose primary has $23.0>i>21.75,$ $21.75>i>20.5,$ and $20.5>i$ respectively. The magnitude separation is used to understand whether contamination is more significant towards faint objects, or in specific regions of the sky -- e.g. by the thick-stream candidates WG1,...,WG4 \citep[][also Shipp et al. in prep.]{agn17} mentioned above.

The fourth column displays their distribution regardless of primary magnitude. The blue points in the last row show the distribution of all WISE-Gaia multiplet targets from Section~2. Among the WISE-DES multiplets with acceptable configurations and colour differences, many have extended morphology\footnote{The DES pipeline assigns every detected object a probability of being an extended source.}, a secondary with $g-i>1,$ and small colour differences ($\left|\delta(g-i)\right|<0.2,$ $\left|\delta(r-z)\right|<0.2$). They can be interpreted as pairs of compact narrow-line galaxies, a common contaminant in quasar lens searches without $u-$band information. 
A na{\"i}ve selection may then require $g-i<0.85$ for all sources. The resulting multiplets with primary $i<20.5$ are shown by blue dots in the second row of Figure~\ref{fig:tank}, and amount to 233 (resp. 179, 405) pairs with secondary $i<20.5$ (resp. $20.5<i<21.75,$ $21.75<i<23.0$). These have been visually inspected, finding some WISE-Gaia candidates independently and yielding at least two good candidates that corresponded to Gaia-DR1 singlets (see tab.~\ref{tab:candtab}, fig.~\ref{fig:multcands}).

\subsection{Future prospects}
This search has mostly relied on a WISE-Gaia preselection, with a final step of DES image inspection. The WISE colour pre-selection, tailored on the majority of known lenses, may nonetheless be too restrictive, as it discards some known lenses (including quads) in the remaining `upper wedge' $W2-W3>\, 3.15+1.5(W1-W2-1.075)$. A crude estimate, based on the number of known lenses that lie in the excluded region, suggests that a further $\approx50\%$ of targets (and possibly candidates) can be identified there.

While quads like \Jxy need little follow-up for confirmation, spatially-resolved spectroscopy of candidate doubles is necessary. A clear example is given by WGD2047-4814 (tab.~\ref{tab:candtab}, fig.~\ref{fig:multcands}): it had already been selected with a different technique \citep{agn15} in DES-Y2 and prioritized as a high-grade candidate, but it was recognized by long-slit spectroscopy (SOAR-Goodman, PI V.~Motta) as a $z=0.33$ narrow-line galaxy aligned with a red galaxy and a blue star (Motta, 2017, private communication). In order to facilitate follow-up, we list the coordinates and $i$-band magnitudes of some high-rank candidates.

A key component of this investigation was the detection of Gaia multiplets, not flagged as duplicate detections, within a $6^\ase$ search radius from WISE-selected quasars. Given the Gaia scanning strategy and object-detection pipeline, some actual multiplets may have been flagged as duplicates in the Gaia-DR1 catalogue. In fact, only about $20-30\%$ of known lenses are currently recognized as multiple unique sources in Gaia-DR1. Future releases may alleviate this issue, since each object is visited multiple times and with different orientations. A complementary search, examining WISE-DES multiplets, may refine the distance/colour cuts explored above (Sect.~\ref{sect:desmul}), adopting a population-mixture approach for pairs/multiplets of objects. This may, in fact, be possible already at the level of Gaia colours, as Gaia-DR2 is also providing `blue' and `red' internal magnitudes.

\section{Summary}
\label{sec:sum}

We have presented the first results of a new search for lensed quasars based on the combination of mid-infrared photometry from WISE, high resolution images from Gaia, and multiband optical imaging from DES. This search has found already-known lenses and some new ones discovered during the STRIDES-2016 campaigns\footnote{Treu et al., and accompanying discovery papers, currently in prep.}, and it yielded $\approx50$ quasar lens candidates in the DES-Y3 footprint, of which 18 high-grade candidates are presented in this paper (tab.~\ref{tab:candtab}, fig.\ref{fig:multcands}).

Out of these, \Jxy is a quadruply lensed quasar, not previously found with other techniques. Its deflector is a massive ($\log(M_{\star}/M_{\odot})\approx11.5$) and low-redshift ($z_{l}\approx0.23$) luminous red galaxy, with a compact bulge, a bright halo contributing most of the light, and indication of isophotal shape variations at large distances.
 Simple lens models can be well fit to its image configuration. The deflector shape parameters obtained from lensing are in general agreement with those observed from the DES $grizY$ cutouts, but there are significant quadrupole degeneracies that will need follow-up imaging to be resolved. The Einstein radius is quite robustly constrained at $(1.30\pm0.04)^\ase,$ whereas the main `halo' component of the deflector has half-light radius $R_{\rm eff}=(3.45\pm0.12)^\ase.$


The double \Jxytwo has been followed up spectroscopically and confirmed as a lens. Its configuration has an arrival-time minimum (image A) well separated from the blend of arrival-time saddle-point (image B) and deflector. A high quadrupole component is expected from our first lens models, and the Einstein radius ($1.3^{\ase}>\theta_{E}>1.0^{\ase}$) can vary appreciably because of this.

The WISE-Gaia-DES multiplet search can be further developed in multiple ways. Multi-band information from DES can be used to augment the multiplet detection, by pruning different classes of contaminants. The WISE colour pre-selection can be performed with looser cuts, or within a population-mixture classification to exclude the most abundant contaminant classes. Furthermore, chromatic information will be available since Gaia-DR2 (due apr.1st 2018), and spectra of most Gaia detections will be available upon completion (2020).  In fact, \Jxy already has a (spatially unresolved) spectrum in the 6dFGS, where the image quality of older imaging surveys precluded its identification as a lensed quasar. 

With this work, we have demonstrated the effectiveness of the WISE-Gaia lens search, discovering new lenses (among which a quadruple), that could not be found with previously developed techniques. At this stage, however, the quantification of how different search techniques are complementary is premature, as comprehensive samples with spectroscopic follow-up are needed to that purpose. Our results suggest that a significant population of lensed quasars may still be found in current surveys, enabling the collection of lens samples for studies of distant quasars (both their hosts and central engines), luminous and dark matter in galaxies over a range of redshifts, and cosmography.

\section*{Acknowledgments}

This paper was written as part of the STRong lensing Insights into the Dark Energy Survey (STRIDES) collaboration, a broad external collaboration of the Dark Energy Survey,
 \texttt{http://strides.astro.ucla.edu}

TT acknowledges support from NSF through grant AST-1450141, and from the Packard Foundation through a Packard Research Fellowship. CDF acknowledges support from the US National Science Foundation through grant number AST-1312329.

Funding for the DES Projects has been provided by the DOE and NSF(USA), MISE(Spain), STFC(UK), HEFCE(UK). NCSA(UIUC), KICP(U. Chicago), CCAPP(Ohio State),  MIFPA(Texas A\&M), CNPQ, FAPERJ, FINEP (Brazil), MINECO(Spain), DFG(Germany) and the Collaborating Institutions in the Dark Energy Survey.  The Collaborating Institutions are Argonne Lab, UC Santa Cruz, University of Cambridge, CIEMAT-Madrid, University of Chicago, University College London,  DES-Brazil Consortium, University of Edinburgh, ETH Z{\"u}rich, Fermilab, University of Illinois, ICE (IEEC-CSIC), IFAE Barcelona, Lawrence Berkeley Lab,  LMU M{\"u}nchen and the associated Excellence Cluster Universe, University of Michigan, NOAO, University of Nottingham, Ohio State University, University of  Pennsylvania, University of Portsmouth, SLAC National Lab, Stanford University, University of Sussex, and Texas A\&M University.  The DES Data Management System is supported by the NSF under Grant Number AST-1138766. The DES participants from Spanish institutions are partially  supported by MINECO under grants AYA2012-39559, ESP2013-48274, FPA2013-47986, and Centro de Excelencia Severo Ochoa SEV-2012-0234. Research leading  to these results has received funding from the ERC under the EU's 7$^{\rm th}$ Framework Programme including grants ERC 240672, 291329 and 306478.

This research has made use of the NASA/ IPAC Infrared Science Archive, which is operated by the Jet Propulsion Laboratory, California Institute of Technology, under contract with the National Aeronautics and Space Administration.

This work has made use of data from the European Space Agency (ESA) mission Gaia (\texttt{https://www.cosmos.esa.int/gaia}), processed by the Gaia Data Processing and Analysis Consortium (DPAC, \texttt{https://www.cosmos.esa.int/web/gaia/dpac/consortium}). Funding for the DPAC has been provided by national institutions, in particular the institutions participating in the Gaia Multilateral Agreement.

This work is based in part on observations obtained at the Southern Astrophysical Research (SOAR) telescope, which is a joint project of the Minist\'{e}rio da Ci\^{e}ncia, Tecnologia, e Inova\c{c}\~{a}o (MCTI) da Rep\'{u}blica Federativa do Brasil, the U.S. National Optical Astronomy Observatory (NOAO), the University of North Carolina at Chapel Hill (UNC), and Michigan State University (MSU).

AA wishes to thank the ITC-Harvard for hospitality in February and June 2017, when most of the work reported here was made.

\section*{Affiliations}
$^1$\eso\\
  $^2$\fnal\\
  $^3$\abello\\
  $^4$\mit\\
  $^5$\ucla\\
  $^6$\valpo\\
  $^7$\eth\\
  $^8$\ioa\\
  $^9$\epfl\\
  $^{10}$\ucd\\
  $^{11}$\ipmu\\
  $^{12}$\kipac\\
  $^{13}$\kavli\\
  $^{14}$\mpa\\
  $^{15}$\swin\\
  $^{16}$\nick\\
  $^{17}$\ctio\\
  $^{18}$\ucl\\
  $^{19}$\rhodes\\
  $^{20}$\cnrs\\
  $^{21}$\sorb\\
  $^{22}$\carn\\
  $^{23}$\stanf\\
  $^{24}$\braza\\
  $^{25}$\brazb\\
  $^{26}$\ifae\\
  $^{27}$\penns\\
  $^{28}$\hyder\\
  $^{29}$\jpl\\
  $^{30}$\calt\\
  $^{31}$\espd\\
  $^{32}$\espa\\
  $^{33}$\mich\\
  $^{34}$\micha\\
  $^{35}$\slac\\
  $^{36}$\illa\\
  $^{37}$\illb\\
  $^{38}$\ohioa\\
  $^{39}$\ohiob\\
  $^{40}$\wash\\
  $^{41}$\aao\\
  $^{42}$\brazc\\
  $^{43}$\brazd\\
  $^{44}$\espc\\
  $^{45}$\south\\
  $^{46}$\camp\\
  $^{47}$\oak\\

\label{lastpage}

\begin{thebibliography}{}
\bibitem[Abazajian et al.(2009)]{aba09} Abazajian, K.~N., Adelman-McCarthy, J.~K., Ag\"{u}eros, M.~A., et al.\ 2009, \apjs, 182, 543
\bibitem[Agnello et al.(2015a)]{agn15} Agnello, A., Kelly, B.~C., Treu, T., \& Marshall, P.~J.\ 2015, \mnras, 448, 1446 
\bibitem[Agnello et al.(2015b)]{agn15b} Agnello, A., Treu, T., Ostrovski, F., et al.\ 2015, \mnras, 454, 1260 
\bibitem[Agnello(2017)]{agn17} Agnello, A.\ 2017, \mnras~ in press, arXiv:1705.08900 
\bibitem[Anguita et al.(2008)]{ang08} Anguita, T., Faure, C., Yonehara, A., et al.\ 2008, \aap, 481, 615 
\bibitem[Bate et al.(2011)]{bat11} Bate, N.~F., Floyd, D.~J.~E., Webster, R.~L., \& Wyithe, J.~S.~B.\ 2011, \apj, 731, 71 
\bibitem[Berghea et al (2017)]{ber17} Berghea, C.~T., et al., arxiv:1705.08359
\bibitem[Chambers et al.(2016)]{cha16} Chambers, K.~C., Magnier, E.~A., Metcalfe, N., et al.\ 2016, arXiv:1612.05560 
\bibitem[Dalal \& Kochanek(2002)]{dal02} Dalal, N., \& Kochanek, C.~S.\ 2002, \apj, 572, 25 
\bibitem[Dark Energy Survey Collaboration et al.(2016)]{des16} Dark Energy Survey Collaboration, Abbott, T., Abdalla, F.~B., et al.\ 2016, \mnras, 460, 1270
\bibitem[Diehl et al.(2016)]{die16} Diehl, H.~T., Neilsen, E., Gruendl, R., et al.\ 2016, \procspie, 9910, 99101D 
\bibitem[Diehl et al.(2017)]{die17} Diehl, H.~T., Buckley-Geer, E., Lindgren, K., et al.\ 2017 ApJS, 232, 15
\bibitem[Ding et al.(2017)]{din17} Ding, X., Liao, K., Treu, T., et al.\ 2017, \mnras, 465, 4634 
\bibitem[Gaia Collaboration et al.(2016)]{gai16} Gaia Collaboration, Prusti, T., de Bruijne, J.~H.~J., et al.\ 2016, \aap, 595, A1 
\bibitem[Gilman et al.(2017)]{gil17} Gilman, D., Agnello, A., Treu, T., Keeton, C.~R., \& Nierenberg, A.~M.\ 2017, \mnras, 467, 3970 
\bibitem[Jones et al.(2004)]{hea04} Jones, D.~H., Saunders, W., Colless, M., et al.\ 2004, \mnras, 355, 747 
\bibitem[Jones et al.(2009)]{hea09} Jones, D.~H., Read, M.~A., Saunders, W., et al.\ 2009, \mnras, 399, 683 
\bibitem[Hutsem{\'e}kers et al.(2015)]{hut15} Hutsem{\'e}kers, D., Sluse, D., Braibant, L., \& Anguita, T.\ 2015, \aap, 584, A61 
\bibitem[Hsueh et al.(2017)]{hsu17} Hsueh, J.-W., Oldham, L., Spingola, C., et al.\ 2017, \mnras, 469, 3713 
\bibitem[Inada et al.(2012)]{ina12} Inada, N., Oguri, M., Shin, M.-S., et al.\ 2012, \aj, 143, 119
\bibitem[King et al.(1999)]{kin99} King, L.~J., Browne, I.~W.~A., Marlow, D.~R., Patnaik, A.~R., \& Wilkinson, P.~N.\ 1999, \mnras, 307, 225 
\bibitem[Kriek et al.(2009)]{kri09} Kriek, M., van Dokkum, P.~G., Labb{\'e}, I., et al.\ 2009, \apj, 700, 221 
\bibitem[van Leeuwen et al.(2017)]{vLe17} van Leeuwen, F., Evans, D.~W., De Angeli, F., et al.\ 2017, \aap, 599, A32
\bibitem[Lin et al.(2017)]{lin17} Lin, H., Buckley-Geer, E., Agnello, A., et al.\ 2017, \apjl, 838, L15 
\bibitem[Lindegren et al.(2016)]{lin16} Lindegren, L., Lammers, U., Bastian, U., et al.\ 2016, \aap, 595, A4 
\bibitem[Mao \& Schneider(1998)]{mao98} Mao, S., \& Schneider, P.\ 1998, \mnras, 295, 587 
\bibitem[More et al.(2016)]{mor16} More, A., Oguri, M., Kayo, I., et al.\ 2016, \mnras, 456, 1595 
\bibitem[Morgan et al.(2004)]{mor04} Morgan, N.~D., Caldwell, J.~A.~R., Schechter, P.~L., et al.\ 2004, \aj, 127, 2617
\bibitem[Motta et al.(2017)]{mot17} Motta, V., Mediavilla, E., Rojas, K., et al.\ 2017, \apj, 835, 132  
\bibitem[Myers et al.(2003)]{mye03} Myers, S.~T., Jackson, N.~J., Browne, I.~W.~A., et al.\ 2003, \mnras, 341, 1 
\bibitem[Oguri et al.(2006)]{ogu06} Oguri, M., Inada, N.,  Pindor, B., et al.\ 2006, \aj, 132, 999 
\bibitem[Oguri et al.(2014)]{ogu14} Oguri, M., Rusu, C.~E., \& Falco, E.~E.\ 2014, \mnras, 439, 2494 
\bibitem[Peng et al.(2006)]{pen06} Peng, C.~Y., Impey, C.~D., Rix, H.-W., et al.\ 2006, \apj, 649, 616 
\bibitem[Rahman et al.(2015)]{rah15} Rahman, M., M{\'e}nard, B., Scranton, R., Schmidt, S.~J., \& Morrison, C.~B.\ 2015, \mnras, 447, 3500 
\bibitem[Rahman et al.(2016)]{rah16} Rahman, M., Mendez, A.~J., M{\'e}nard, B., et al.\ 2016, \mnras, 460, 163 
\bibitem[Refsdal(1964)]{ref64} Refsdal, S.\ 1964, \mnras, 128, 307 
\bibitem[S{\'a}nchez \& DES Collaboration(2010)]{san10} S{\'a}nchez, E., \& DES Collaboration 2010, Journal of Physics Conference Series, 259, 012080 
\bibitem[Schechter et al.(2014)]{sch14} Schechter, P.~L., Pooley, D., Blackburne, J.~A., \& Wambsganss, J.\ 2014, \apj, 793, 96 
\bibitem[Schechter et al.(2017)]{sch17} Schechter, P.~L., Morgan, N.~D., Chehade, B., et al.\ 2017, \aj, 153, 219 
\bibitem[S{\'e}rsic(1968)]{ser68} Sersic, J.~L.\ 1968, Cordoba, Argentina: Observatorio Astronomico, 1968,  
\bibitem[Shanks et al.(2015)]{sha15} Shanks, T., Metcalfe, N., Chehade, B., et al.\ 2015, \mnras, 451, 4238 
\bibitem[Sluse et al.(2012)]{slu12} Sluse, D., Hutsem{\'e}kers, D., Courbin, F., Meylan, G., \& Wambsganss, J.\ 2012, \aap, 544, A62 
\bibitem[Suyu et al.(2017)]{suy17} Suyu, S.~H., Bonvin, V., Courbin, F., et al.\ 2017, \mnras, 468, 2590
\bibitem[Williams et al.(2017)]{wil17} Williams, P., Agnello, A., \& Treu, T.\ 2017, \mnras, 466, 3088 
\bibitem[Wright et al.(2010)]{wri10} Wright, E.~L., Eisenhardt, P.~R.~M., Mainzer, A.~K., et al.\ 2010, \aj, 140, 1868-1881 
\bibitem[Yonehara et al.(1999)]{yon99} Yonehara, A., Mineshige, S., Fukue, J., Umemura, M., \& Turner, E.~L.\ 1999, \aap, 343, 41 
\end{thebibliography}
\end{document}